\documentclass[useAMS,usenatbib]{mn2e}
\input epsf
\usepackage{graphicx}
\sloppypar

\title[High mass X-ray binaries in the SMC: the luminosity function]{High mass X-ray binaries in the SMC: the luminosity function}
\author[P. Shtykovskiy and M. Gilfanov]{P. Shtykovskiy$^{1,2}$
\thanks{E-mail: pav\_sht@hea.iki.rssi.ru; gilfanov@mpa-garching.mpg.de} 
and M. Gilfanov$^{2,1}$\footnotemark[1]\\
$^{1}$Space Research Institute, Russian Academy of Sciences, 
      Profsoyuznaya 84/32, 117997 Moscow, Russia\\
$^{2}$Max-Planck-Institute f\"ur Astrophysik, 
      Karl-Schwarzschild-Str. 1, D-85740 Garching bei M\"unchen, 
      Germany}
\begin{document}

\date{Accepted. Received; in original form}

\pagerange{\pageref{firstpage}--\pageref{lastpage}} \pubyear{2005}

\maketitle

\label{firstpage}

\begin{abstract}
We study population of compact X-ray sources in the Small Magellanic
Cloud using the archival data of XMM-Newton observatory. The total
area of the survey is $\approx 1.5$ square degrees with the limiting
sensitivity of $\approx 10^{-14}$ erg/s/cm$^2$, corresponding to the
luminosity of $\approx 4.3\cdot 10^{33}$ erg/s at the SMC distance. 
Out of $\sim 150$ point sources detected in the 2--8 keV energy band,
$\sim3/4$  are background CXB sources, observed
through the SMC. Based on the properties of the optical and near-infrared
counterparts of the detected  sources we identified  likely HMXB
candidates, and  sources, whose nature is uncertain, thus, providing
a lower and upper limits on the luminosity distribution of HMXBs
in the observed part of the SMC. 

The observed number of HMXBs is consistent with the prediction based on 
SFR estimates derived from the supernovae frequency and analysis
of color-magnitude diagrams of the stellar population.
If, on the contrary, the true  value of the SFR  is better
represented by FIR, H$_{\alpha}$ and UV based estimators, then the
abundance of HMXBs in the SMC may significantly (by a factor of as
much as $\sim$10) exceed the value derived for the Milky Way and other
nearby galaxies. 
The shape of the observed distribution at the bright end is consistent
with the universal HMXB XLF. At the faint end, $L_X\la 2\cdot 10^{34}$
erg/sec, the upper limit on the luminosity function is consistent with 
while the lower limit is significantly flatter than the $L^{-0.6}$
power law.   
\end{abstract}


\section{Introduction}
\label{sec:intro}

X-ray observations of the Small Magellanic Cloud (SMC) revealed a surprisingly rich
population of High Mass X-ray Binaries in this close neighbour of the
Milky Way \citep[e.g.][]{asca,BeXraySMC}.
At the time of writing, about 50 sources of this type have been
identified in the SMC with the majority of discovered systems being Be/X-ray
binaries. Existence of such a rich population of HMXBs in a nearby galaxy
opens unique possibility to study  properties of population of HMXBs.

As has been shown by \citet{grimm03}, the X-ray luminosity function
(XLF) of HMXBs obeys, to the first approximation, a universal power
law distribution with the differential slope of $\approx 1.6$,
whose normalization is proportional to the star formation rate of the
host galaxy. The validity of this universal HMXB XLF has been established
in a broad range of the star formation rates and regimes and in the
luminosity range   $\log(L_X)\ga 35.5-36$. Based on the ASCA
observations of the Small Magellanic Cloud and on the behavior of the
integrated X-ray luminosity of distant galaxies located at redshifts
$z\sim 0.3-1.3$ observed by Chandra in the Hubble Deep Field North,
\citet{grimm03} tentatively suggested, that the HMXB XLF is not
dramatically affected by the metallicity variations.

Study of the population of high mass
X-ray binaries in the SMC is of importance for several reasons:
\begin{enumerate}
\item
The Magellanic Clouds are known to have a significant
under-abundance of metals \citep{mc_book97}. For example, the interstellar medium
of the SMC has a mean metallicity 0.6 dex lower than the local ISM,
the [Fe/H] of young objects (age$<0.6$Gyr) in the SMC is 0.5~dex lower
than in the solar vicinity \citep{mc_book97}. 
The effects of the metallicity
variations on the population of X-ray binaries are poorly understood. 
Study of the SMC, a galaxy with relatively well known chemical composition
gives a unique opportunity to investigate such effects
observationally. 
\item
Owing to the proximity of the SMC, the weakest sources become reachable
within a moderate observing time.
Indeed, the sensitivity of a typical Chandra or XMM-Newton observation,
$\sim 10^{-14}$ erg/s/cm$^2$ corresponds to the luminosity of 
$\sim 4.3\cdot 10^{33}$ erg/s at the SMC distance. 
This opens the possibility to study the low luminosity end of the
HMXB XLF.
In addition, the proximity of the SMC allows to study the properties 
of its stellar population and the star formation history in details.
\end{enumerate}

\begin{table*}
 \centering
 \begin{minipage}{140mm}
  \caption{List of XMM-Newton observations used for analysis.}
\label{tb:pnt}
  \begin{tabular}{@{}lccccc@{}}
  \hline
Obs. ID   & Target & R.A.  & Dec.  & Instrument & Exposure\\
          &        & J2000 & J2000 &            & ksec \\
  \hline
0135721501  &1ES0102-72       &  16.0297  &-72.0109 & PN &25ks  \\
0112880901&  CF Tuc &13.3742 &-74.6614  & MOS1+MOS2  & 80ks \\
0110000201& IKT 18     &14.9441 & -72.1583 & MOS1+MOS2  &39ks  \\
0110000301&IKT 23     &16.3021 &-72.3744  & MOS1+MOS2  &58ks  \\
0110000101&IKT 5      &12.3645 &-73.2215 & MOS1+MOS2  &53ks  \\
0084200101&SMC Pointing 1  & 14.1015 & -72.3590 & MOS1+MOS2  & 33ks  \\
0084200801&SMC Pointing 8   &13.7263 & -73.6701 & MOS1+MOS2  & 42ks \\
0011450101&SMC X-1    &19.1750 &-73.4381 & MOS1+MOS2  &102ks  \\
0157960201&XTE J0055-727   &13.9263 &-72.7143  & MOS2  & 19ks  \\
\hline
\end{tabular}
\end{minipage}
\end{table*}

In the present paper we study population of X-ray binaries in the SMC
based on the archival XMM-Newton data.
The distance modulus of the SMC is  $m-M=18.9$
\citep{mc_book97}, corresponding to the distance of $D\approx 60$
kpc.

The paper is structured as follows. XMM-Newton observations and data
analysis are described  in the section ~\ref{sec:data}. 
In the section ~\ref{sec:nature} we discuss nature
of detected X-ray sources. In the section
~\ref{sec:ident} we describe the HMXBs search procedure and its
results. Resulting HMXBs luminosity function is 
presented in the section ~\ref{sec:hmxbpop}. 
The log(N)--log(S) of CXB in the direction of the SMC 
is presented in the section ~\ref{sec:cxbsrc}. 

\begin{figure}
\includegraphics[width=0.5\textwidth,clip=true]{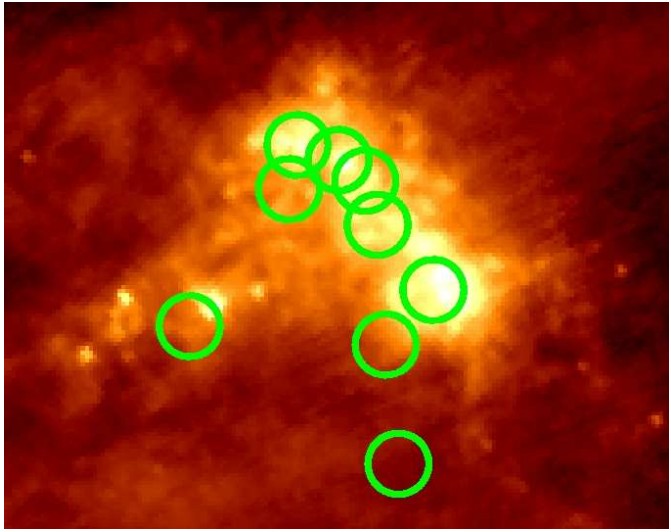}
\caption{Far-infrared (IRAS, $100\mu$) map of the  region around the
Small Magellanic Cloud. 
Circles show fields of view of the XMM--Newton observations used for
the analysis.  
}
\label{fig:fir_map}
\end{figure}

\section{Observations and data analysis}
\label{sec:data}

We have selected 9 XMM-Newton archival observations with the pointing
direction towards the SMC and with sensitivity better than
$\sim {\rm few}\times 10^{-14}$ erg/s/cm$^2$ in the 2--10
keV energy band. These observations are listed in Table
\ref{tb:pnt}. Fig.\ref{fig:fir_map} shows their fields of view,
overlayed on the far-infrared map (IRAS, $100\mu$) of the Small
Magellanic Cloud. 

The observations were processed with Science Analysis System (SAS) v6.0.0. 
After filtering out high background intervals we extracted
images in the 2--8 keV energy band. This energy range was chosen to
minimize the fraction of  cataclysmic variables
and foreground stars among detected sources, generally having softer
spectra than high mass X-ray binaries. To improve the sensitivity of the 
survey, images from MOS1 and MOS2 detectors have been merged. 
If both MOS and PN data were available, we used the data having 
higher sensitivity. This was determined comparing the exposure times
and taking into account that the effective area of PN
is $\approx$ two times higher than the effective area of each of MOS
cameras.

\subsection{Source detection}

The source detection  was performed with
standard SAS tasks {\em eboxdetect} and {\em emldetect}. 
The value of the threshold likelihood $L$ used in emldetect task to
accept or reject detected source was chosen as follows.
We simulated a number of images with Poisson background (without
sources). Each of the generated images was analyzed with the full sequence
of the source detection procedures using different values of the
emldetect threshold likelihood $L$ and for each trial value of $L$ the
number of detected ``sources'' was counted. The final value of the
threshold likelihood, $L=10.5$, was chosen such 
that the total number of spurious detections was $\la 1$ per 9
images.

The obtained images  were visually inspected and  the source lists were
manually filtered of spurious sources near bright sources and
 sources whose extent undoubtedly exceeded PSF size. 
The final merged source list contains $\approx 150$ sources.
Their flux distribution is plotted in Fig.\ref{fig:lf0}.

\subsection{Energy conversion factor}

The 2--8 keV source counts were converted to the 2--10
keV energy flux assuming a power law spectrum with the photon index
1 and N$_H$=5$\cdot10^{20}$cm$^{-2}$. 
The energy conversion factors are 
ecf$_{\rm MOS}$=2.95$\cdot10^{-11}$ erg/cm$^2$ and 
ecf$_{\rm PN}$=1.0$\cdot10^{-11}$ erg/cm$^2$.
We note that, as will be shown in sec.~\ref{sec:cxb}, significant part of 
sources in our sample are background AGN.
Therefore, in sections  ~\ref{sec:cxb} and ~\ref{sec:cxbsrc}
 (figures ~\ref{fig:area}, ~\ref{fig:lf0} and ~\ref{fig:cxb})
 we used energy conversion factors calculated for photon index 1.7, 
ecf$_{\rm MOS}$=2.27$\cdot10^{-11}$ erg/cm$^2$ and 
ecf$_{\rm PN}$=8.0$\cdot10^{-12}$ erg/cm$^2$, instead 
of values mentioned above.

\subsection{Boresight correction}

For several XMM observations we performed boresight correction
using optical sources from the MCPS catalogue \citep{mcps}. 
The correction was applied in case when two or 
more well-known sufficiently bright X-ray sources 
with optical counterparts were found.

\subsection{Correction for incompleteness}
\label{sec:area}

Naturally,  the point source detection sensitivity varies
from observation to observation and, for a given observation, across
the telescope field of view. The reasons of these variations are
difference in the exposure times, deterioration of the point spread
function of the telescope with the increase of the off-axis angle,
presence of the diffuse emission and background variations.
These factors affect the completeness of the survey at low fluxes. 
The conventional way to account for these effects without setting too high
conservative completeness limit, is to calculate
 (using the sensitivity maps of individual observations) the survey area as a 
function of flux.
After appropriate normalisation it represents the fraction 
of sources we are able to detect
as a function of flux, and can be used 
to correct the detected number of low-flux sources:
\begin{equation}
\left(\frac{dN}{dS}\right)_{corrected}=\frac{A_0}{A(S)}\cdot\left(\frac{dN}{dS}\right)_{observed}
\label{eq:lc1}
\end{equation}
where A(S) is the survey area at flux S and A$_0$ is the total (geometrical) 
survey area.
The corrected cumulative log(N)--log(S) distribution  can
be obtained as follows: 
\begin{equation}
N(>S)=\sum_{S_j>S}{\frac{A_0}{A(S_j)}} ,
\label{eq:lc}
\end{equation}
where $S_j$ is the flux of the j-th source.

\begin{figure}
\includegraphics[width=0.5\textwidth]{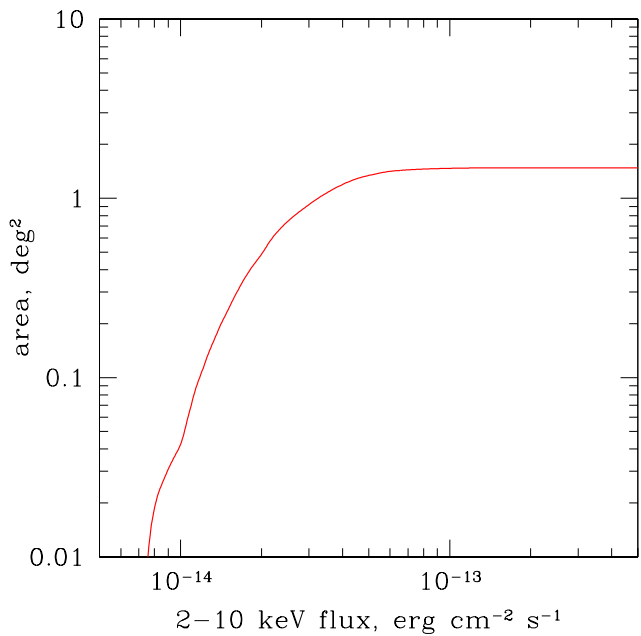}
\caption{The survey area as a function of 2--10 keV flux, calculated
as described in section \ref{sec:area}.}
\label{fig:area}
\end{figure}

The flux-dependent survey area calculated as described in
\citet{lmcpaper} is shown in Fig.~\ref{fig:area}. 
From Fig.~\ref{fig:area} one can see that the incompleteness effects
become important at fluxes  $\la10^{-13}$erg/s/cm$^2$. In the high
flux limit, the total area of the survey equals $A_{\rm tot}\approx
1.48$ deg$^2$.
 The flux distribution,
corrected for the incompleteness effects, is plotted as the thick
histogram in Fig.\ref{fig:lf0}.

\section{Nature of X-ray sources in the field of the SMC}
\label{sec:nature}

\subsection{Background and foreground sources}
\label{sec:cxb}

\begin{figure}
\includegraphics[width=0.5\textwidth]{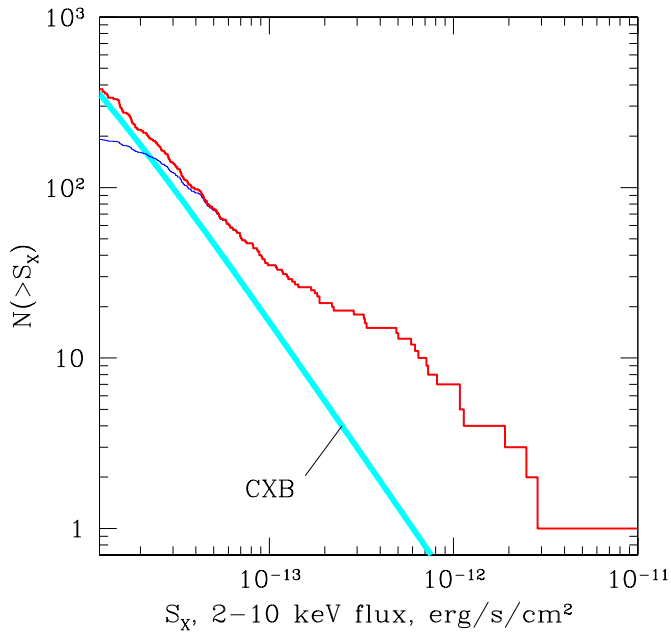}
\caption{Cumulative log(N)--log(S) distribution of detected point-like  
sources, excluding several  known foreground stars and rotation
powered pulsars (section \ref{sec:nature}). 
The thin and thick histograms show, respectively, the
observed distribution and the distribution corrected for the
incompleteness effects, as described in section \ref{sec:area}.  
The thick  grey line shows the log(N)--log(S) of CXB sources according  
to \citet{cxb}.
The brightest source SMC X-1 with flux S$_X=6.15\cdot10^{-10}$erg/s/cm$^2$
is situated far outside the flux range of the plot.
}
\label{fig:lf0}
\end{figure}

The total number of sources with the flux 
$F_X[2-10 {\rm ~keV}]>2.3\cdot 10^{-14}$ erg/s/cm$^2$ (corresponding to the luminosity 10$^{34}$erg/s at the SMC distance) is
151. With account for the survey  incompleteness (Eq.~\ref{eq:lc})
this number corresponds to 
$N(>2.3\cdot 10^{-14})\approx 192$ 
(see thick histogram in Fig.~\ref{fig:lf0}). 
According to the CXB
$\log(N)-\log(S)$ determined by \citet{cxb}, the total number of CXB
sources expected in the field of 1.48 deg$^2$ is 
$N_{CXB}\approx 148$. From the comparison of these numbers it is
obvious that about 3/4 of sources in our sample are background AGNs.

A significantly less important source of contamination is X-ray
sources associated with foreground stars in the Galaxy (no known
Galactic X-ray binaries were located in the field of view of XMM
observations).
Foreground stars are removed by our filtering procedure, as described
in section~\ref{sec:ident}.

\subsection{Low mass X-ray binaries}
\label{sec:lmxb}

Given the limiting sensitivity of the survey, $F_X\sim (1-3)\cdot
10^{-14}$ erg/s/cm$^2$, corresponding to the luminosity $L_X\sim
4.3\cdot10^{33}-1.3\cdot10^{34}$ erg/s at the SMC distance, the intrinsic SMC
sources are dominated by X-ray binaries. Their total number is
proportional to the stellar mass (LMXB) and star formation rate
(HMXB) of the galaxy. 

The total stellar mass of the SMC can be estimated from the integrated
optical luminosity. According to RC3 catalog \citep{rc3}, the
reddening corrected V-band magnitude  of the SMC equals 
$V_{To}\approx 1.92$, corresponding to the total
V-band luminosity of $L_V\approx 4.6\cdot10^{8}$ $L_{\sun}$. From the
dereddened optical color $(B-V)_{To}\approx 0.36$ \citep{rc3} and using
results of \citet{m2l}, the V-band mass-to-light ratio is
$(M/L)_V\approx 0.59$ in solar units, giving the total stellar mass of the
SMC  $M_*\approx 2.7\cdot10^{8}$M$_{\sun}$. 
For this stellar mass,
using results of \citet{gilfanov04}, about one LMXB with
luminosity $L_X\ga 10^{35}$ erg/s is expected in the entire galaxy.    
Such a prediction agrees with observations as no
 LMXB candidates are known in the SMC at present.

\subsection{Star formation rate in the SMC}
\label{sec:sfr}
Determination of a star formation rate in the SMC, a diffuse galaxy
with low opacity is a nontrivial task.

Below we compare estimates of the star formation rate in the SMC,
obtained from different SFR indicators.
The SFR values refer to the star formation rate of stars in the
0.1-100 M$_{\sun}$ range, assuming Salpeter's IMF.
The discussion of results of different star formation indicators 
could be also found in \citet{wilke04} 
(see also discussion of the SFR in the LMC by \citet{lmcpaper}).

\citet{wilke04} based on ISOPHOT observations and calibration
of \citet{kenn98} obtained the star formation rate in the SMC 
SFR(IR)=0.015$~M_\odot/$yr.
\citet{wilke04}, however, noted that this value underestimates the
SFR of the SMC because some stellar
radiation, due to low opacity of the SMC, can escape the star formation 
regions directly, and gave a more realistic estimate of
 \begin{eqnarray}
{\rm SFR(IR)}=0.05~M_\odot/{\rm yr}.
\label{eq:sfr_ir1}
\end{eqnarray}
With the total infrared luminosity of the SMC \citet{wilke04}
L$_{IR}=3.21\cdot10^{41}$erg/s and using the equation (5) from \citet{bell03}
which accounts for the uncertainty mentioned above, we obtained
 \begin{eqnarray}
{\rm SFR(IR)}=0.044~M_\odot/{\rm yr}.
\label{eq:sfr_ir2}
\end{eqnarray}
Note, that this calibration should be applied extremely
cautious for such a low-luminosity galaxy \citep{bell03}.

The H$_{\alpha}$ luminosity-based indicators can also underestimate
the star formation rate in the SMC due to the reasons mentioned above.
Based on H$_{\alpha}$ luminosity of the SMC and applying extinction
correction \citet{kenn95} estimated the star formation rate of 
 \begin{eqnarray}
{\rm SFR(H_{\alpha})}=0.044~M_\odot/{\rm yr}.
\label{eq:sfr_halpha1}
\end{eqnarray}
\citet{kenn91} gave an conservative upper limit of
 \begin{eqnarray}
{\rm SFR(H_{\alpha})}\la0.1~M_\odot/{\rm yr}.
\label{eq:sfr_halpha2}
\end{eqnarray}

An SFR estimator relatively free of this uncertainty is the one based
on combined UV and FIR emission.
With the UV flux measured by the D 2B-Aura satellite
\citep{uv80} at $\lambda=1690\AA$
 corrected for foreground extinction
and uncorrected ISOPHOT FIR flux we obtain with the
\citet{kenn98} calibration
 \begin{eqnarray}
{\rm SFR(UV+FIR)}=0.063~M_\odot/{\rm yr}.
\label{eq:sfr_uvir}
\end{eqnarray}
This value is consistent with the SFR(UV) derived from flux
corrected for both foreground and internal extinction \citep{uv80}: 
\begin{eqnarray}
{\rm SFR(UV,~ext.corr.)}=0.063~M_\odot/{yr} 
\label{eq:sfr_uv}
\end{eqnarray}

\citet{filipovic98}, from comparison of discrete radio (Parkes
telescope) and X-ray (ROSAT) sources, estimated a number of SNRs in
the SMC (14). 
From the age -- radio flux density relation
they estimated an SNR birth rate of one SNR in $350\pm70$ yrs.
We convert this value to the SFR in the mass range 0.1-100M$_{\sun}$, assuming
Salpeter's IMF and taking into account that only stars 
with M$\geq8$~M$_{\sun}$ can result in a radio SNR:
\begin{eqnarray}
{\rm SFR(SNR~ birth~ rate)}\approx0.38^{+0.10}_{-0.05}~M_\odot/{\rm yr}.
\label{eq:sfr_snr}
\end{eqnarray}

Similar value has been obtained by \citet{sfhistory} who
studied star formation history of the SMC
fitting the color-magnitude diagrams of stellar population with
theoretical isochrones: 
\begin{eqnarray}
{\rm SFR(CMD)}\approx0.37~M_\odot/{\rm yr}.
\label{eq:sfr_cmd}
\end{eqnarray}
This number corresponds to recent star formation which took place
$\sim4-5$Myr ago. 

Thus, the SFR values derived using  different star formation rate
proxies  can differ by the factor of as much as $\sim$ten.
The ``standart''  indicators (IR, UV, H$_{\alpha}$) result in the
rather low value of $\sim0.05$ M$_{\sun}$/yr, whereas 
the estimates based on the radio supernova rate and the analysis of
the color-magnitude diagram point towards much 
larger star formation rates, in the $\sim0.3-0.4$ M$_{\sun}$/yr range. 
We note, that  the methods based on the analysis of the
color-magnitude diagrams are generally believed to be the most
reliable and accurate. 
As the controversy between different SFR indicators can not be
resolved at present, we shall  assume in the following that the true
value of SFR is somewhere in the range:  
\begin{eqnarray}
{\rm SFR(SMC)}\approx 0.05-0.4~M_\odot/{\rm yr}.
\label{eq:sfr}
\end{eqnarray}

\subsection{High mass X-ray binaries}
\label{sec:hmxb}

For the star formation rate of SFR=$0.05-0.4~M_{\sun}$ yr$^{-1}$, 
using  HMXB-SFR relation of
\citet{grimm03} \citep[see also the discussion of the normalization in
  the HMXB-SFR
  calibration by ][]{lmcpaper}
we predict from $\sim6$ to $\sim49$ HMXBs with luminosities $\geq10^{35}$erg
s$^{-1}$ in the whole SMC. This confirms that the population of X-ray
binaries in the SMC is dominated by HMXBs.

To predict the number of HMXBs in our sample of X-ray sources, we
estimate the star formation rate in the part of the SMC covered by XMM 
pointings from IRAS infrared maps provided by {\em SkyView} \citep{skyview}. 
Calculating the far infrared flux according to the formula
FIR$=1.26\cdot 10^{-11}(2.58S_{60\mu}+S_{100\mu})$, where FIR is flux
in erg/s/cm$^2$ and $S_\lambda$ is flux in Jy 
\citep{helou85}, and integrating the IRAS maps we obtained the total
SFR$=0.011M_{\sun}$/yr. This number is a factor of $\sim 4.5\div36.4$ smaller
than eq.(\ref{eq:sfr}) because of the uncertainties mentioned in
sec.~\ref{sec:sfr}.  
To make it consistent
with our determination of the SFR in the SMC (eq.\ref{eq:sfr}), we
simply multiply the FIR-based values  by the correction factor of
$\approx 4.5\div36.4$. The thus calculated star formation rate of the
part of the 
SMC covered by XMM observations, is: 
\begin{eqnarray}
{\rm SFR(XMM)}\approx 0.018\div0.15M_\odot/{\rm yr}.
\label{eq:sfr_xmm}
\end{eqnarray}

With the above value of SFR we predict about $\sim2-18$ HMXBs
with luminosities 
$\geq10^{35}$ erg s$^{-1}$ in the observed part of the SMC. The error in
this number accounts for uncertainty of the SFR estimate and does
not include the Poisson fluctuations.

\subsection{The Log(N)--Log(S) distribution}

As demonstrated above, the population of compact X--ray sources in the
SMC field is dominated by two types of sources -- background AGN and
high mass X-ray binaries in the SMC.  
Their log(N)--log(S) distributions  in the flux range of interest can
be described by a power law 
with the differential slopes $\approx 2.5$ (CXB) and 
$\approx 1.6$ (HMXBs). Due to significant difference in the
slopes, their relative contributions depend strongly on the 
flux. At large fluxes, $F_X\ga (2-3)\cdot 10^{-13}$ erg/s/cm$^2$ 
($L_X\ga 10^{35}$ erg/s) the X-ray
binaries in the SMC prevail. On the contrary, in the low 
flux limit, e.g. near the sensitivity limit of our survey, $F_X\sim
10^{-14}$ erg/s/cm$^2$, the majority ($\sim3/4$) of the X-ray sources are
background AGN.

This is illustrated by Fig.~\ref{fig:lf0}, showing the observed and
corrected for incompleteness  log(N)--log(S) distribution of all
sources from the final source list. The corrected log(N)--log(S)
distribution agrees at low fluxes with that of CXB sources. At high
fluxes, there is an apparent excess of sources above the numbers
predicted by the CXB sources log(N)--log(S), due to the contribution
of HMXBs.

\section{Identification of HMXB candidates}
\label{sec:ident}

To filter out contaminating background and foreground sources, we use
the same filtering procedure as described by \citet{lmcpaper}.
It is based on the fact that optical emission from HMXBs is
dominated by their optical companions, whose properties, such as 
absolute magnitudes and intrinsic colors are
sufficiently well known.  

\subsection{Optical properties of HMXBs counterparts}
\label{sec:hmxbopt}

High mass X-ray binaries  are powered by accretion of mass
lost from the massive early-type optical companion. 
The mechanism of accretion could be connected either to (i) strong
stellar wind from an OB supergiant (or bright giant) or (ii)
equatorial circumstellar disk around Oe or Be type star
\citep[e.g.][]{corbet, jvp95}.
In the case of Small Magellanic Cloud where the only supergiant system
SMC X-1 is known at present time, the HMXBs with Be companion seem to give the
main contribution \citep{hmxbopt}.

Given the large number of well known high mass X-ray binaries
in the SMC, the selection criteria for the optical counterparts
of such systems can be determined straightforwardly.
For example, \citet{hmxbopt} studied the 
optical properties of Be/X-ray pulsars in the SMC.
The majority of HMXBs with identified optical companions in 
their sample (32 out of 34) have
apparent V-band magnitudes m$_V=14-17$.
The exceptions are supergiant system SMC X-1 with m$_V=13.2$
and X-ray pulsar CXOU J010042.8--721132 with m$_V=18.01$
 \citep[OGLE catalogue][]{ogle} which has been 
 previously classified by \citet{axp} as a possible anomalous X-ray pulsar (AXP).

This picture is consistent with the expectations based on the 
positions of possible optical counterparts on the
Hertzsprung-Russel diagram, distance modulus of the SMC, 
$(m-M)_0=18.9$ \citep{mc_book97}, and foreground and intrinsic
extinction towards the SMC, A$_V\sim0.46$ (for hot population)
\citep{mcps}.
OB supergiants and bright giants (luminosity classes I--II)  have
absolute magnitudes M$_V\sim -7 - -4$, corresponding to apparent 
magnitudes in the range m$_V\sim12.5-15.5$.
The position of Be stars in the H--R diagram
is close to the main sequence. 
Apparent magnitudes m$_V<17-18$ correspond to spectral classes 
earlier than B3--B5 for main sequence stars (B5--B7 for giants).
Thus we can conclude, that HMXBs in the SMC with main sequence Be 
optical companions have spectral classes earlier than B3--B5 
(B5--B7 for giants), in agreement with expectations based on the optical 
properties of HMXBs in our Galaxy.

As potential optical counterparts of HMXBs belong to spectral classes
earlier than $\sim$ B3--B5, their intrinsic optical and near-infrared
colors are constrained by $B-V\la-0.20$, $J-K\la-0.16$. With account for the
interstellar reddening, both apparent colors are expected to be $\la 0.1-0.2$. 
This is consistent with B-V colors of HMXBs
from \citet{hmxbopt} sample. In infrared part of the spectrum, however, 
Be stars are known to exhibit
an infrared continuum excess, caused by free-free and free-bound emission 
within the disk \citep{beinfrared}.
Therefore, J--K colors of Be stars can be higher than values qouted above.
To avoid this uncertainty, 
in identification procedure we use mainly B--V colors, applying
J--K identification criteria only in extreme cases 
(in fact we used this criteria alone only for one source).

\subsection{Catalogs and selection criteria}
\label{sec:id_proc}

We used the following optical and near-infrared catalogs:
\begin{enumerate}
\item Magellanic Clouds Photometric Survey: the SMC  (MCPS) \citep{mcps}
\item Guide Star Catalog, version 2.2.1 (GSC2.2.1) \citep{gsc}
\item The CCD survey of the Magellanic Clouds \citep{massey}
\item 2-micron All Sky Survey (2MASS) \citep{2mass}
\item Emission-line stars and PNe in the SMC \citep{meyssonnier}
\item A 2dF survey of the Small Magellanic Cloud \citep{2df}
\end{enumerate}

As a first step, we cross-correlated the XMM-Newton X-ray source list
with MCPS, GSC and \citet{massey} optical catalogs, using search
radius 4$\arcsec$ and the following selection criteria: 
$12.0<V<18.0$,  $B-V<0.6$.
If optical object was present in MCPS and either GSC or \citet{massey} 
catalogs, the
preference was given to MCPS, as it has a higher photometric accuracy
(especially at low magnitudes). 
The color limits are higher than possible colors
of HMXBs optical companions (B--V$\la$0.1-0.2) and were chosen to account for
limited photometric accuracy of optical catalogues. 
We found  optical counterparts for 54 X-ray sources using these 
selection criteria.

On the second stage, we cross-correlated all X-ray sources having
optical counterparts with near-infrared 
catalog 2MASS, catalog of \citet{meyssonnier} and catalog of \citet{2df}  
using the same search radius as before.

Then the following filtering procedure was applied.
\begin{enumerate}
\item 
In case when source was present only 
in the GSC catalogue (which has rather low photometric accuracy), 
we applied filtering criteria based on the near infrared data.
In fact, there was only one source with J--K$\approx$1.2 rejected by
this criteria.

\item All X-ray sources with low X-ray-to-optical flux ratio 
$F_X/F_{\rm opt}<10^{-3}$  were rejected. The optical flux was
calculated from the  V-band magnitude,
F$_{\rm opt}=8.0\cdot10^{-6}\cdot10^{-m_{V}/2.5}$ erg/s/cm$^{2}$.
Such low $F_X/F_{\rm opt}<10^{-3}$ ratios are typical for foreground
stars but not for X-ray binaries. 
All confirmed HMXBs in our sample
have $F_X/F_{\rm opt}>3\cdot10^{-3}$ among which $\ga90\%$ have 
$F_X/F_{\rm opt}>10^{-2}$.

\end{enumerate}

\subsection{Search radius}
\label{sec:searchrad}
The choice of the search radius plays a crucial role 
in the procedure of search for potential HMXB candidates based on
their optical companions.
Too small value of the search radius may lead to incompleteness of resulting
HMXBs list (especially at low luminosities), while 
too large value may lead to a significant fraction of chance coincidences. 
Given the spatial density of optical stars (whose optical colors and
magnitudes satisfy our selection criteria) in the MCPS catalogue,
the number of detected X-ray sources and the search radius of 4$\arcsec$,
we estimate the number of chance coincidences to be $\approx14$.
On the other hand, a typical error on the position of a
low-luminosity source is $\sigma\approx2\arcsec$ (1$\sigma$ confidence).
Therefore, in ideal situation with the search radius of 4$\arcsec$ we miss 
only $\sim13\%$ of low-luminosity sources.
The real fraction of missed optical counterparts, however, could be somewhat
higher as it was obtained assuming ideal astrometric accuracy.
Our choice of the search radius is further justified in the Appendix.

\subsection{Identification results}
\label{sec:id_res}
\label{sec:indsrc}

After the filtering, 50 X-ray sources remained in the list of
potential HMXB candidates (Table \ref{tb:hmxbcand}). 
As follows from sec.\ref{sec:searchrad}, $\approx14$ of these sources can be
 chance coincidences.
Therefore, we further split the filtered list into reliable HMXBs
and the sources whose nature is less certain. We include in the list of 
reliable HMXB candidates 
only sources which have either detected X-ray pulsations or
optical emission lines, or sources whose spectral class is well-determined based
on their optical spectrum.
The latter two identification criteria are mainly based on \citet{meyssonnier} and
\citet{2df} catalogues.

There were 32 sources satisfying at least one of these criteria. 
Remaining 18 sources are considered to be sources of
uncertain nature.

The resulting likely HMXBs candidates and sources of uncertain nature
are listed in Table~\ref{tb:hmxbcand}.  
All known  HMXBs located in the field of view of the XMM-Newton observations are
among these sources. 

Comments on the individual sources:

\#1: To minimize the influence of pile-up on the flux of SMC X-1, for this source
we used data from EPIC PN camera operating in Small Window mode.

\#17 = RXJ0049.2--7311: Optical colors consistent with HMXB nature.
This source was identified with an bright H$_{\alpha}$ 
object by \citet{hmxbopt}
and is probably associated with nearby X-ray pulsar AXJ0049-732 (p=9s). 

\#34: Optical colors consistent with HMXB nature, however, infrared 
colors are relatively high, J--K=0.97.

\#35: Optical colors consistent with HMXB nature.
\citet{axp} classified this source as
a possible anomalous X-ray pulsar (p=8s).

\#38: This source is situated inside the NGC 330 stellar cluster.
As a stellar density is high there, we 
consider the photometry to be unreliable. 

\#49: Optical colors consistent with HMXB nature.
This source is situated in 24$\arcsec$ from source AXJ0048.2--7309.
\citet{yoko03} based on the presence of emission-line object in \citet{meyssonnier}
catalogue within the ASCA error circle of 40$\arcsec$ ($90\%$ confidence) 
and X-ray spectrum classified this ASCA source as a possible Be X-ray pulsar.
 The posititon of the XMM source is inconsistent with emission-line
object in \citet{meyssonnier} and therefore either optical identification of ASCA
source was incorrect either XMM source is unrelated to ASCA source.
Therefore we consider this source as a source of uncertain nature.

\begin{table*}
\caption{List of HMXB candidates}
\label{tb:hmxbcand}
\begin{centering}
\begin{tabular}{lcccccccl}
\hline
\#  & R.A.  & Dec.  &   $L_X$ (1) &$m_V$&$B-V$ & F$_x$/F$_{opt}$& id &Comments       \\
 & & &    erg s$^{-1}$ &   &     \\     
\hline
\multicolumn{8}{c}{\em Likely HMXB candidates}\\
\hline
1  & 01 17 05.2& -73 26 38   &$3.8\cdot10^{38}$ &13.2 &-0.14
&0.55 & p, ob&SMC X-1 P=0.717s (2) \\

2  & 00 47 23.3& -73 12 28   &$1.4\cdot10^{36}$ &16.0 &0.08
&0.80 & p & RX(AX)J0047.3--7312 P=263s (2)  \\

3  & 00 51 52.3& -73 10 34   &$1.1\cdot10^{36}$ &14.5 &-0.07 & 0.14
& p,em & RXJ0051.9--7311=AXJ0051.6--7311 P=172s (2) \\

4  & 00 49 42.0& -73 23 14   &$6.4\cdot10^{35}$ &14.9 &0.15 &0.12&p,em
& RXJ0049.7--7323=AXJ0049.5--7323 P=755.5s (2)\\

5  & 00 57 50.4& -72 07 55   &$6.1\cdot10^{35}$ &15.7 &-0.05
&0.26&p,em & CXOUJ005750.3--720756 P=152.3s (2) \\

6  & 00 54 55.9& -72 45 10   &$5.2\cdot10^{35}$ &15.0 &-0.03 &0.14&p,em &AXJ0054.8--7244 P=500s(2,3) \\

7  & 00 57 49.6& -72 02 35   &$4.6\cdot10^{35}$ &15.7 &
-0.12&0.19&p,em & RXJ0057.8--7202=AXJ0058-72.0 P=280.4s (2) \\

8  & 01 01 20.8& -72 11 17    &$4.0\cdot10^{35}$ &15.6 &-0.19 &0.15
&p,em& RXJ0101.3-7211 P=455s (2) \\

9  & 01 01 52.1& -72 23 34  &$4.0\cdot10^{35}$ &14.9 &-0.01 &0.084
&em, ob& AXJ0101.8--7223(2) \\

10 & 00 55 18.5& -72 38 52    &$3.6\cdot10^{35}$ &15.9 &
0.15 &0.18 &p& P=702s (3)  \\

11 & 00 59 20.9& -72 23 15  &$3.5\cdot10^{35}$ &15.0 &-0.01 &0.075
&p&XMMUJ005921.0--722317 P=202s (4) \\

12 & 00 50 57.3& -73 10 09    &$3.3\cdot10^{35}$ &14.4 &0.08 &0.040
&em& RXJ0050.9--7310=AXJ0050.8--7310 (2) \\

13 & 00 50 44.6& -73 16 05  &$2.9\cdot10^{35}$ &15.5 &-0.11
&0.097&p,em &RXJ0050.8--7316 AXJ0051--733 P=323.2 (2) \\

14 & 01 03 13.4& -72 09 13    &$2.7\cdot10^{35}$ &14.8 &-0.06 &0.053
&p,em& SAXJ0103.2--7209=AXJ0103.2--7209 P=345.2 (2)  \\

15 & 00 58 12.4& -72 30 50  &$2.0\cdot10^{35}$ &14.9 &0.11 &0.038
&em$^{5}$ &RXJ0058.2--7231 (2,5) \\

16 & 00 55 28.5& -72 10 57 &$1.8\cdot10^{35}$ &16.8 &-0.12 &0.21 &p &
CXOUJ005527.9--721058 P=34s (6)\\

17 & 00 49 13.7& -73 11 38 &$1.3\cdot10^{35}$ &16.4 &0.19 &0.10& em & =(?)AXJ0049--732(P=9s) RXJ0049.2--7311 (2,7,10)\\

18 & 00 48 34.3& -73 02 31  &$9.9\cdot10^{34}$ &14.8 &0.0 &0.023 &em, ob&
RXJ0048.5-7302 (2)  \\

19 & 00 54 56.4& -72 26 48   &$8.1\cdot10^{34}$ &15.3 &-0.05 &0.023
&p,em& XTEJ0055--724 P=59s (2)  \\

20 & 01 05 55.3& -72 03 51   &$6.6\cdot10^{34}$ &15.7 &-0.06 &0.028
&em &RXJ0105.9--7203 AXJ0105.8--7203 (2) \\

21 & 01 01 37.4& -72 04 18  &$6.1\cdot10^{34}$ &16.3 &-0.16 &0.044 &em
& RXJ0101.6--7204 (2) \\

22 & 00 56 15.2& -72 37 54   &$4.9\cdot10^{34}$ &14.6 &0.13 &0.0075
&em, ob&  \\

23 & 00 57 36.1& -72 19 33  &$4.8\cdot10^{34}$ &16.0 &0.01 &0.027&em,p &CXOUJ005736.2--721934 P=565s (2)\\

24 & 00 57 24.0& -72 23 57  &$4.0\cdot10^{34}$ &14.7 & -0.07&0.0068 & ob& \\

25 & 00 49 29.9& -73 10 58  &$3.5\cdot10^{34}$ &16.2 &0.2 &0.023 &em& =(?)AXJ0049--732(P=9s)(2,7) \\

26 & 01 05 08.1& -72 11 48  &$3.5\cdot10^{34}$ &15.7 &-0.08 &0.016 &em
&RXJ0105.1--7211 AXJ0105--722 P=3.34s? (2)\\

27 & 01 03 37.7& -72 01 35  &$3.4\cdot10^{34}$ &14.7 &-0.11 &0.0055
&em& RXJ0103.6--7201 (2)  \\

28 & 01 19 39.0& -73 30 14   &$3.3\cdot10^{34}$ &15.9 &-0.07 &0.016 &em& (4) \\

29 & 01 01 03.0& -72 07 01  &$2.7\cdot10^{34}$ &15.8 &-0.07 &0.013
&em&RXJ0101.0--7206 P=304.5 (2) \\

30 & 01 00 30.3& -72 20 32    &$2.1\cdot10^{34}$ &14.6 &-0.06 &0.0034
&em, ob& XMMUJ010030.2--722035 (2)  \\

31 & 00 56 05.8& -72 21 57   &$2.0\cdot10^{34}$ &15.9 & -0.04&0.097
&p,em& XMMUJ005605.2--722200 P=140.1 (2)\\

32 & 00 50 47.9& -73 18 12  &$8.7\cdot10^{33}$ &15.6 &0.12 &0.0035 &em&  \\

\hline
\multicolumn{8}{c}{\em Sources of uncertain nature}\\
\hline

33 & 00 55 34.9& -72 29 05  &$2.1\cdot10^{35}$ &14.7 &-0.04 &0.031& &  \\

34 & 00 48 18.8& -73 21 00  &$1.1\cdot10^{35}$ &16.2 &0.25 &0.069 & & \\

35 & 01 00 42.9& -72 11 32 &$1.1\cdot10^{35}$ &17.7 &-0.03 &0.28 &p &
CXOUJ010043.1--721134 AXP? P=8s (8)\\

36 & 00 54 32.2& -72 18 09 &$7.2\cdot10^{34}$ &16.6 &-0.08 &0.067 & & \\

37 & 00 54 03.6& -72 26 30  &$6.6\cdot10^{34}$ &14.9 &0.01 &0.014 & & \\

38 & 00 56 18.7& -72 28 02   &$3.9\cdot10^{34}$ &15.3 &-0.40 &0.012 & & NGC330
stellar cluster\\

39 & 01 00 37.2& -72 13 16  &$2.5\cdot10^{34}$ &16.7 & -0.17&0.026 & & \\

40 & 00 48 33.4& -73 23 55  &$2.5\cdot10^{34}$ &17.7 &0.07 &0.064 & & \\

41 & 00 54 41.1& -72 17 20 &$2.1\cdot10^{34}$ &15.7 & -0.11&0.0090 &  &\\

42 & 00 47 19.9& -73 08 22 &$2.0\cdot10^{34}$ &17.3 &0.08 &0.036 & & \\

43 & 00 53 35.8& -72 34 24 &$1.7\cdot10^{34}$ &17.2 &-0.07 &0.029 & & \\

44 & 01 04 37.4& -72 06 30  &$1.4\cdot10^{34}$ &18.0 &0.13 &0.048 & & \\

45 & 01 00 22.9& -72 11 27  &$1.2\cdot10^{34}$ &17.0 &-0.13 &0.018 &  &\\

46 & 00 48 04.0& -73 17 01  &$1.2\cdot10^{34}$ &17.0 &0.21 &0.016 &  &\\

47 & 00 49 05.4& -73 14 10   &$1.0\cdot10^{34}$ &18.0 &0.03&0.034 & & \\

48 & 01 03 28.3& -72 06 51  &$1.0\cdot10^{34}$ &16.5 &-0.15 &0.0086 & & \\

49 & 00 48 14.9& -73 10 03  &$8.6\cdot10^{33}$ &15.3 &0.26 &0.0025 & &
=(?)AXJ0048.2--7309 (2,9)\\

50 & 00 50 20.2& -73 11 17  &$8.6\cdot10^{33}$ &17.0 & 0.11&0.012 & & \\

\hline


\end{tabular}\\
\end{centering}
\smallskip
(1) -- 2--10 keV band, assuming distance of 60 kpc; 
(2) -- see \citet{BeXraySMC} and references inside;
(3) -- \citet{haberllongperiod};
(4) -- \citet{majid};
(5) -- \citet{edgecoe};
(6) -- \citet{edge};
(7) -- \citet{filipovic00};
(8) -- \citet{axp};
(9) -- \citet{yoko03};
(10) -- \citet{hmxbopt};
Comments: 
p -- source shows pulsations;
em -- source has Balmer emission lines in its optical spectrum
(based on \citet{meyssonnier} catalogue, unless
otherwise is noted);
ob -- source is present in \citet{2df} catalogue and belongs to OB
spectral class;
\end{table*}

\begin{figure}
\includegraphics[width=0.5\textwidth]{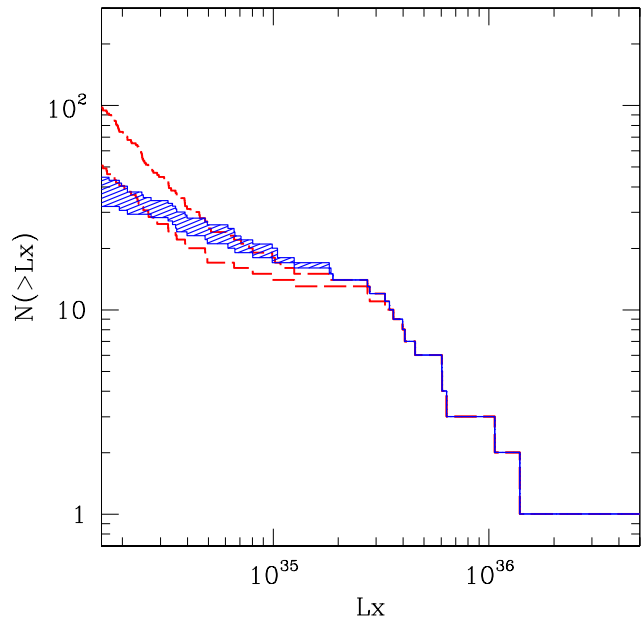}
\caption{The incompleteness-corrected XLFs of HMXB candidates in
the SMC obtained with two different HMXB search methods (see sec.\ref{sec:hardness}). Two solid histograms correspond to upper and lower limits on XLF
obtained with search method based on the optical properties of HMXBs.
Two dashed histograms correspond to upper and lower limits on XLF
obtained with search method based on the X-ray hardness ratio.
It is clear from this figure, that both methods give similar result
for sources with luminosities $L_X\ga2-3\cdot10^{34}$erg/s.
The brightest source SMC X-1 with luminosity L$=3.8\cdot10^{38}$erg/s
is situated outside the luminosity range of the graph.
}
\label{fig:hardnessfilt}
\end{figure}

\begin{figure}
\includegraphics[width=0.5\textwidth]{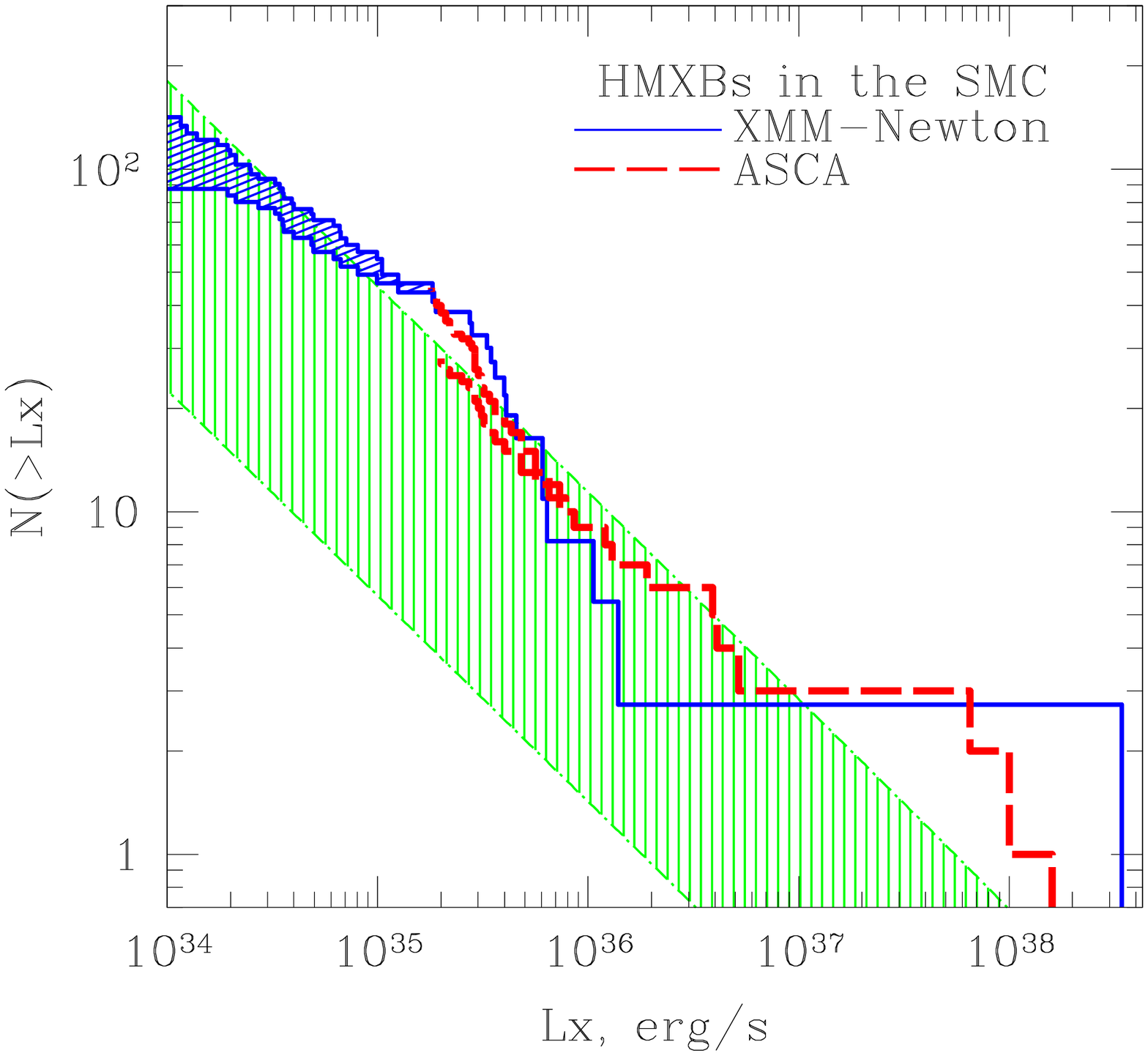}
\caption{
The upper and lower limits on the  XLF of HMXBs in the SMC. 
The solid histogram shows XMM-Newton results scaled up by the factor
of  ${\rm SFR_{tot}/SFR_{XMM}}\approx 2.7$ according to the 
contribution of the XMM-Newton survey area to the total SFR of the SMC.
The dashed histogram is the ASCA XLF from \citet{yoko03}.
The shaded area shows the luminosity distribution  
predicted from the ``universal'' HMXB XLF of \citet{grimm03}
extrapolated towards low luminosities. Its width reflects the
uncertainty of our knowledge of the SFR in the SMC, as 
discussed in the section \ref{sec:sfr}.
}
\label{fig:lfhmxbxmmasca}
\end{figure}

\subsection{Completeness}
\label{sec:optcompleteness}

The completeness of the list of HMXB candidates is defined by the
following factors: 

\begin{enumerate}
\item The completeness of the optical catalogues.
The initial search for optical counterparts is  based on
MCPS and GSC2.2 catalogs. 
In general, the MCPS catalogue is thought to be complete for V$<$20 \citep{mcps}.
The GSC2.2 is a magnitude selected ($V\le19.5$) subset of the  GSC-II catalog 
(http://www-gsss.stsci.edu/gsc/gsc2/GSC2home.htm). The latter is 
complete to $J=21$ at high galactic latitudes \citep{gsc}. 

Completeness of both catalogs is known to break down in the crowded regions.
As no sensitivity maps for the optical catalogs exist, a quantitative
estimate of the completeness of the initial counterpart search
is impossible.
However, the quoted completeness limits of both catalogs are
$2-3$ mag better than  chosen threshold of $18$ mag for the
optical counterparts search. This suggests that the completeness of
the  optical catalogs is unlikely to be the primary limiting factor.  

The MCPS catalogue also has ``stripe-like'' gaps with zero stellar density.
We checked that the majority of X-ray sources do not coincide
with this ``gaps''.
For few sources which were close or inside the gaps we used two additional
catalogs: OGLE \citep{ogle} and (if OGLE had no data in this region) 
USNO-B \citep{usno-b}.

\item The efficiency of the initial search due to statistical and
systematic uncertainties in the positions of X-ray and optical
sources. This is probably one of the major limiting factors.
As has been noted in sec~\ref{sec:searchrad}, the search radius of 4$\arcsec$
assuming ideal astrometric quality corresponds to 
detection of $\sim87\%$ of the weakest sources, while for moderate
luminosity and bright sources this fraction would be close to 100$\%$.
In Appendix we show, that in real case
(i.e. with all uncertainties taken into account), 
with the chosen value of the search radius we miss not more
than few sources.

\item The filtering procedure applied to the optical matches found in
the initial search. This procedure is based mainly on observed
optical and near-infrared properties of known HMXBs in the SMC and Milky Way.
The distribution of V-band magnitudes and B-V optical colors of
HMXBs in the SMC have maxima far inside our filtering limits 
(see Fig.~\ref{fig:opticalproperties}), meaning that we detect the vast
majority of HMXBs unless some significantly different (and previously
undetected) type of these systems exists.
\end{enumerate}

Thus, we estimate the completeness of our HMXBs sample to be
very high with the uncertainties in the positions of X-ray sources
being the main factor of incompleteness.
We note, that as this factor is flux-dependent, it affects
most strongly the low-luminosity part of our sample.

\subsection{Comparison with other HMXB identification methods}
\label{sec:hardness}

The identification method based on the properties of optical
counterparts is not the only way to identify the population of HMXBs
in the background of contaminating sources.
For example, \citet{asca} found that hardness ratios (HRs) of all known
X-ray binary pulsars
(XBPs) in the Magellanic Clouds fall into narrow ``XBP region'', 
HR$_{XBP}$(ASCA)$=0.2-0.6$.
In order to compare these two methods, we apply selection criteria based on hardness ratio 
to our sample of X-ray sources.

We use XBPs selection criteria similar to those described in \citet{asca}.
The hardness ratio is defined as $HR=(H-S)/(H+S)$, where H and S are
count rates in the 2-7 keV and 0.7-2 keV energy bands respectively.
All X-ray sources with HRs corresponding to photon indexes 
 $\alpha=0.4-1.5$ (corresponding ``XBP region'' defined by \citet{asca}) 
are classified as X-ray binary pulsar candidates.
X-ray sources which have slightly softer spectra, $\alpha=1.5-2.0$, 
are considered as sources of uncertain nature.
Higher photon indexes are untypical for HMXBs, especially for Be/X-ray
binaries which seem to constitute the majority of HMXBs population in the SMC.
Luminosity distributions obtained using these two methods are shown 
in Fig.~\ref{fig:hardnessfilt}.

As is clear from Fig.~\ref{fig:hardnessfilt}, the results of
both methods agree for sources with L$_X\ga2-3\cdot10^{35}$erg/s, while
for the low-luminosity sources the method based on hardness ratio gives
significantly higher number of sources.
This could be a result of large statistical uncertainty in the hardness ratios
for low flux sources, leading to a ``leakage'' of background AGN to the XBP
region. In addition, the population of highly absorbed AGN with truly hard
spectra becomes important at low fluxes and compromises the simple
hardness ratio based selection criteria.

\section{Population of HMXBs in the SMC}
\label{sec:hmxbpop}

\subsection{The luminosity function of HMXBs  in the SMC}
\label{sec:hmxbcandlf}

The incompleteness-corrected luminosity distribution of HMXBs
in the SMC is shown in Fig.~\ref{fig:lfhmxbxmmasca}. In order to represent
the HMXB population of the entire galaxy, the observed XLF has been
scaled up according to the relative contribution of the area covered  
by the XMM-Newton survey to the total SFR of the SMC.
The upper and lower histograms below $\sim 2\cdot 10^{35}$ erg/sec
correspond to all sources from Table \ref{tb:hmxbcand} and 
to the likely HMXB candidates respectively. These two histograms
provide upper and lower limits on the true X-ray luminosity function
of HMXBs in the observed part of the SMC. 

Fig.~\ref{fig:lfhmxbxmmasca} also shows the XLF of 
HMXB candidates in the SMC obtained by \citet{yoko03} based on
the ASCA data. The ASCA XLF refers to the entire SMC. We note that
removal of the sources located  outside XMM pointings does not change
its shape significantly. The upper ASCA histogram in the figure
corresponds to all sources from  \citet{yoko03} excluding SNR, AGN and   
foreground stars -- i.e. represents all sources which properties do
not contradict to HMXB nature. The lower histogram shows the
luminosity distribution of likely HMXB candidates -- confirmed  X-ray 
binary pulsars and candidates and confirmed non-pulsating HMXBs and
candidates (see \citealt{yoko03} for details). 
As is evident from the figure, the XMM-Newton and ASCA XLFs are
consistent with each other. The K-S test for two distribution gives
the  probability of $\sim50-70\%$. 

\subsection{Abundance of HMXBs in the SMC}

The shaded area in Fig.~\ref{fig:lfhmxbxmmasca}  shows the universal HMXB
XLF of \citet{grimm03} and its extrapolation towards low
luminosities. The width of the shaded area reflects uncertainties in
our knowledge of the star formation rate in the SMC as discussed in the
section \ref{sec:sfr}. As is clear from Fig.~\ref{fig:lfhmxbxmmasca},
the observed abundance of HMXBs in the SMC is consistent with the upper
range of the predicted numbers, which is based on the SFR estimates
derived from the  supernovae rate and analysis of the stellar
color-magnitude diagram.  
If, on the contrary, the true value of the SFR is better represented
by FIR, H$_{\alpha}$ and UV based estimators, then the abundance of
HMXBs in the SMC may exceed significantly, by a factor as much as
$\sim$10, the value derived for the Milky Way and other nearby galaxies
by \citet{grimm03}. Note that the analysis and interpretation of the
SMC data  in \citet{grimm03} was based on the star formation rate
estimates obtained by \citet{filipovic98} from the supernovae rate.

\begin{figure}
\hbox{
\includegraphics[width=0.5\textwidth]{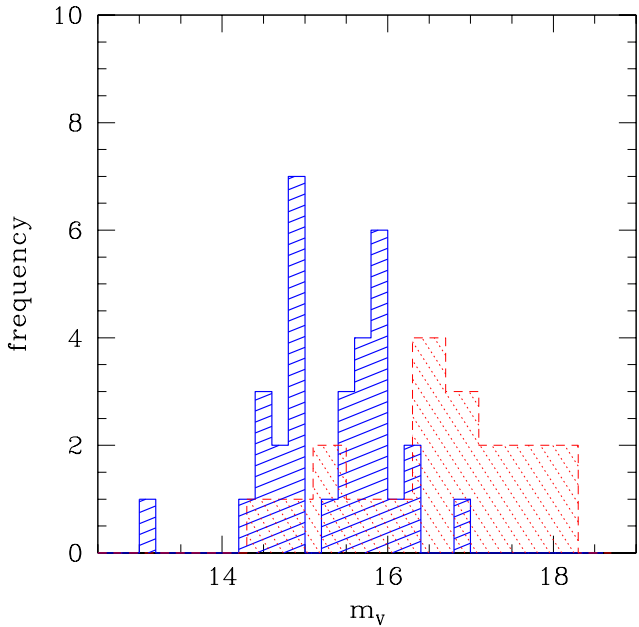}
}
\caption{
Distribution of the V-band magnitudes of the HMXB candidates in the
SMC.
The thick histogram corresponds to the likely HMXB candidates from the
upper part of the table~\ref{tb:hmxbcand}.
The thin histogram corresponds to the uncertain sources from the lower
part of the table~\ref{tb:hmxbcand}.
}
\label{fig:opticalproperties}
\end{figure}

\subsection{The XLF shape}

At the first glance the shape of the HMXB  XLF obtained by XMM-Newton
appears to be more complex than a single slope $N(>L)\propto L^{-0.6}$
power law in the entire $\sim 10^{34}-10^{38}$ erg/sec luminosity
range. 

The maximum likelihood fit to the bright end of the distribution
($L_X>2\cdot 10^{35}$erg/s) with a power law  results in the best fit
value of the cumulative slope $\alpha=0.74^{+0.22}_{-0.19}$ which is
consistent with the slope of 0.6. This is also confirmed by the
Kolmogorov-Smirnov (K-S) test which gives  probability of $\sim17\%$. 
Thus, the bright end of the XLF is consistent  with the universal HMXB
XLF. This conclusion is further confirmed by the ASCA XLF containing
by a factor of $\sim 3$ more sources.

At low luminosities the uncertainties of HMXB identification
procedure become of importance and we consider the upper and lower
limits of the XLF separately. 
The upper limit on the XLF (the upper histogram
in Fig.~\ref{fig:lfhmxbxmmasca})   is consistent with a single slope
power law in the luminosity range L$_X>10^{34}$ erg/s with the K-S
test probability of $\sim 75\%$ and  the best fit value of the slope
$\alpha=0.5\pm0.08$.  
For the lower histogram in Fig.~\ref{fig:lfhmxbxmmasca} (luminosity
distribution of reliable HMXB candidates) we obtain best fit slope
of $\alpha=0.37\pm0.08$ in the full luminosity range. 
The observed distribution, is consistent with a single slope power law 
-- the K-S test probability is $\approx 33\%$. 
A power law approximation in the $10^{34}-2\cdot
10^{35}$ erg/sec luminosity range gives the best fit slope of 
$\alpha\approx 0.13_{-0.13}^{+0.30}$.

Thus, the behavior of the XLF in the faint end is somewhat
uncertain. However, based on the statistical arguments and the
distribution of the optical magnitudes (see sec.~\ref{sec:searchrad}
and ~\ref{sec:hmxboptobserved}), the majority of sources of
uncertain nature are expected to be the result of chance coincidence. 
Therefore the lower histogram in Fig.~\ref{fig:lfhmxbxmmasca} should
better represent the true HMXB XLF and the flattening of
the XLF in the faint end can be real. It can result from the
``propeller'' effect, as suggested by \citet{lmcpaper}.

\subsection{Optical properties of HMXB candidates}
\label{sec:hmxboptobserved}
The distribution of V-band magnitudes of the HMXB candidates is
shown in Fig.~\ref{fig:opticalproperties}.
In the case of the likely HMXB candidates it
shows bimodal structure with two peaks at m$_V\approx14.6$ and
m$_V\approx15.5$.
To estimate the statistical significance of the observed bimodality, we 
perform a test based on the bootstrap method \citep{silverman}.
The obtained value of significance is rather low and does not allow
us to make a definite conclusion (with probability that the observed distribution 
is multimodal of 97\%, i.e. significance slightly above 2$\sigma$).

The origin of this bimodality, if real, is not clear. It could
correspond to Be/X-ray systems with main sequence and giant Be secondaries.
However, this explanation is not satisfactory as it would require a rather
narrow distributions of the optical companions over spectral classes.

The distribution of the m$_V$ magnitudes of the uncertain HMXB candidates
differs from those of likely candidates.
This is consistent with expectations, that significant part of them
should be a chance coincidences (see sec.~\ref{sec:searchrad}).

\begin{figure}
\hbox{
\includegraphics[width=0.5\textwidth]{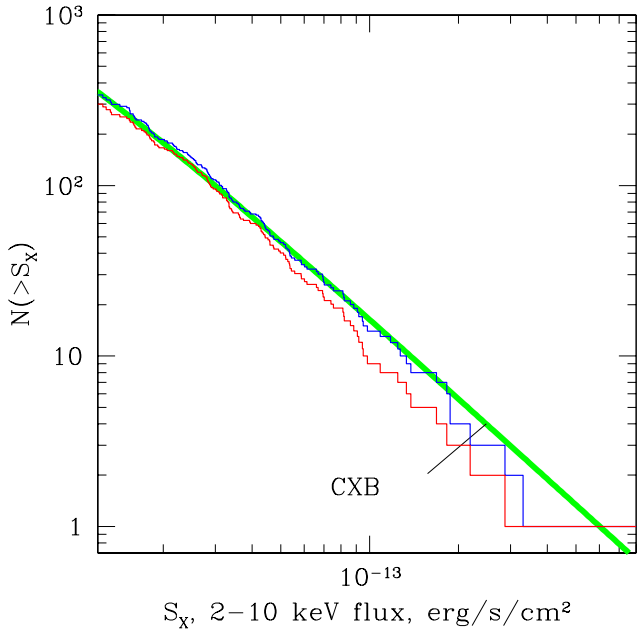}
}
\caption{Cumulative logN--logS distribution of CXB sources, obtained
after removal of all HMXB candidates (lower histogram) and only likely HMXB candidates
(upper histogram). 
The solid line shows the CXB log(N)--log(S) distribution from \citet{cxb}.
The brightest source has flux S$_X=2.85\cdot10^{-12}$erg/s/cm$^2$.
}
\label{fig:cxb}
\end{figure}

\section{Log(N)--log(S) distribution of CXB sources in the direction
of SMC} 
\label{sec:cxbsrc}

The log(N)--log(S) distributions of CXB sources, obtained after removal
of likely HMXB candidates only and all HMXB candidates
are shown in Fig.\ref{fig:cxb}. 
We fit the resulting log(N)--log(S) distribution in the flux range 
$F_X>2\cdot10^{-14}$ erg cm$^{-2}$ s$^{-1}$  with a power law model
$N(>S)=k(S/S_0)^{-\alpha}$, where $S_0=2\cdot10^{-14}$
erg/s/cm$^2$.  
Best fit value of slope for the case with only likely HMXB candidates excluded is 
$\alpha=1.48\pm0.12$. 
For the case with all the HMXB candidates excluded we obtain 
 $\alpha=1.55\pm0.13$. 

We also compare the resulting log(N)--log(S) distributions in the flux 
range $F_X>2\cdot10^{-14}$ erg cm$^{-2}$ s$^{-1}$ with CXB number counts 
obtained by \citet{cxb}.
The Kolmogorov-Smirnov test accepts their model giving probabilities 
$\sim96\%$ for the case with likely HMXB candidates excluded
and $\sim76\%$ for the case with all the HMXB candidates excluded.
The number of sources at flux $2\cdot10^{-14}$ erg cm$^{-2}$ s$^{-1}$ 
predicted by \citet{cxb}, N$=178^{+96}_{-46}$, is also consistent with 
the observed numbers, N$=167\pm13$ and  N$=186\pm14$ for the two above
cases respectively.

Thus, the CXB log(N)--log(S) in the direction of the SMC is consistent
both in the shape and the normalization with results of other CXB
surveys.

\section{Summary}
\label{sec:summary}

Based on the archival data of XMM--Newton observations, we studied the
population of point-like sources in the field of the SMC. The total area of
the survey is $\approx 1.5$ sq.degr. with the limiting sensitivity of
$\sim 10^{-14}$ erg/s/cm$^2$ (Fig.\ref{fig:area}), corresponding to
the luminosity of $\sim 2.3\cdot 10^{33}$ erg/s at the SMC distance. 
\begin{enumerate}
\item
Out of the 196 compact sources, detected in the 2--8 keV energy
band, $\sim3/4$ are CXB sources, observed through
the SMC (section \ref{sec:cxb}, Fig.\ref{fig:lf0}). 
\item
Based on the stellar mass and the star formation rate of the SMC we
demonstrate, that the majority of the intrinsic SMC
sources, detected in the 2--8 keV band are high mass X-ray binaries 
(section \ref{sec:nature}). 
\item
The proximity of the SMC and adequate angular resolution of
XMM--Newton make it possible to reliably filter out the sources,
whose properties are inconsistent with HMXB nature. 
Based on the optical and infrared
magnitudes and colors of the optical counterparts of the X-ray sources
we identify 32 likely HMXB candidates (29 of which were previously
known) and 18 sources of uncertain
nature (section \ref{sec:ident}, Table \ref{tb:hmxbcand}).  
The remaining $\sim 146$ sources, with a few exceptions, are
background objects, constituting the resolved part of CXB. Their
flux distribution is consistent both in shape and normalisation with other determinations of
the  CXB log(N)--log(S) (section \ref{sec:cxbsrc},
Fig.\ref{fig:cxb}). 
\item
With these results we constrain lower and upper bounds of the
luminosity distribution of HMXBs in the observed part of the SMC.
We compare these with  the extrapolation
towards low luminosities of the  universal luminosity function of
HMXBs, derived by \citet{grimm03}  
(Fig.~\ref{fig:lfhmxbxmmasca}). 

The observed number of HMXBs is consistent with the prediction based
on SFR estimates derived from supernovae frequency and analysis of the
stellar color-magnitude diagrams. 
If, on the contrary, the true  value of the SFR  is better
represented by FIR, H$_{\alpha}$ and UV based estimators, then the
abundance of HMXBs in the SMC may significantly (by a factor of as
much as $\sim$10) exceed the value derived for the Milky Way and other
nearby galaxies (sections \ref{sec:sfr}, \ref{sec:hmxbpop}). 

The shape of the observed distribution at the bright end is consistent
with the $N(>L)\propto L^{-0.6}$ power law of the universal HMXB XLF. 
At the faint end, $L_X\la 2\cdot 10^{35}$ erg/s, the upper limit of
the luminosity function is consistent with the   $L^{-0.6}$ power law,
while the lower limit is significantly  flatter. 
\end{enumerate}

\section{Acknowledgements}
This research has made use of data obtained through the High Energy 
Astrophysics Science Archive Research Center Online Service, provided 
by the NASA/Goddard Space Flight Center.
This publication has made use of data products from the Two Micron All
Sky Survey,  Guide Star Catalogue-II. 
PS acknowledges the support of European Association for Research in Astronomy
(MEST-CT-2004-504604 Marie Curie - EARA EST fellowship at 
Max-Planck-Institute for Astrophysics).
PS also acknowledges a partial support from the President of the
Russian Federation grant SS-2083.2003.2.

\appendix
\section{Cross-corellation of two catalogues}
Below we consider the problem of cross-corellation  of two catalogs
(e.g. of the list of detected X-ray sources and an optical catalogue).
More specifically we concentrated on the case when 
only a fraction of X-ray sources has true optical
 counterparts in the optical catalogue and when the probability of 
chance coincidences is not negligible.
In this case, the matches of X-ray sources with optical
 sources will include real optical counterparts and chance coincidences.
We consider the behavior of the growth curves -- the dependence
of the number of matches on the search radius. 
We show that using different dependences of the number of 
true matches and chance coincidences
on the search radius, it is possible to estimate
the number of true optical counterparts.

\subsection{Case of constant density of optical stars}
The number of X-ray sources with optical matches inside
radius r depends on the number of X-ray sources which have true
optical counterpart inside r and number of chance coincidences of
X-ray sources with field stars:
\begin{equation}
N_{Xmatch}=N(1-e^{-\rho\pi r^2})+Me^{-\rho\pi r^2}\int_0^r\phi(r')dr',
\label{eq:appendix1}
\end{equation}
where N - total number of X-ray sources, M - number of X-ray sources
with true optical counterparts, $\rho$ - density of optical field stars
(i.e. excluding the true optical counterparts) and $\phi(r)$
is the distribution of distances  between X-ray sources and their 
optical counterparts.
The first term in the above equation is a number of chance coincidences, i.e. 
 number of X-ray sources which have at least one field star inside its search radius.
The second term corresponds to number of true matches falling inside search radius r.
The exponent in this term excludes cases when both
true optical counterpart and chance coincidence with field stars are present, as
such matches were already taken into account in the first term.
Assuming that all X-ray sources have the same positional error
$\sigma$ (hence $\phi(r)=r/\sigma^2\cdot e^{-r^2/2\sigma^2}$), we can rewrite this equation 
in the following way: 
\begin{equation}
N_{Xmatch}=N(1-e^{-\rho\pi r^2})+Me^{-\rho\pi r^2}(1-e^{-r^2/2\sigma^2}).
\label{eq:appendix2}
\end{equation}
Fitting the curve N$_{Xmatch}$(r) with this formula one can determine the number of X-ray
sources with true optical matches, surface density of optical stars
$\rho$ and error on the source positions.

\subsection{Non-constant surface density of optical stars}
\label{sec:appvardens}

In a real situation, the density of optical stars $\rho$ varies from
field to field and even from source to source.
This can lead to significant deviations of N$_{Xmatch}$(r) curve
behaviour from the  eq.~\ref{eq:appendix2}.
There is no obvious way to overcome this problem completely.
However, the density variations could be partially taken into account
by replacing the mean density of optical objects
in the first term of eq.~\ref{eq:appendix2} by  
the local densities of optical objects around each individual 
X-ray source.
The first term in this case, of course, should be replaced by
the corresponding sum over all X-ray sources.
The density $\rho$ in the second term, however, 
cannot be straightforwardly calculated because 
it is related to X-ray sources with true optical 
counterparts, which we do not know a priori.
Thus, modifying the first term in equation \ref{eq:appendix2} we obtain:
\begin{equation}
N_{Xmatch}=\sum_{i=1}^{N}(1-e^{-\rho_i\pi r^2})+Me^{-\rho\pi r^2}(1-e^{-r^2/2\sigma^2}).
\label{eq:appendix3}
\end{equation}
The local densities $\rho_i$ of optical objects should be
calculated individually for each X-ray
source. 
The areas over which these densities are calculated must be
large enough to minimize the relative contribution of the true optical
counterparts and Poissonian errors.
Density $\rho$ corresponds now to the mean density of optical objects
near X-ray sources which have true optical counterparts.

\begin{figure}
\includegraphics[width=0.45\textwidth]{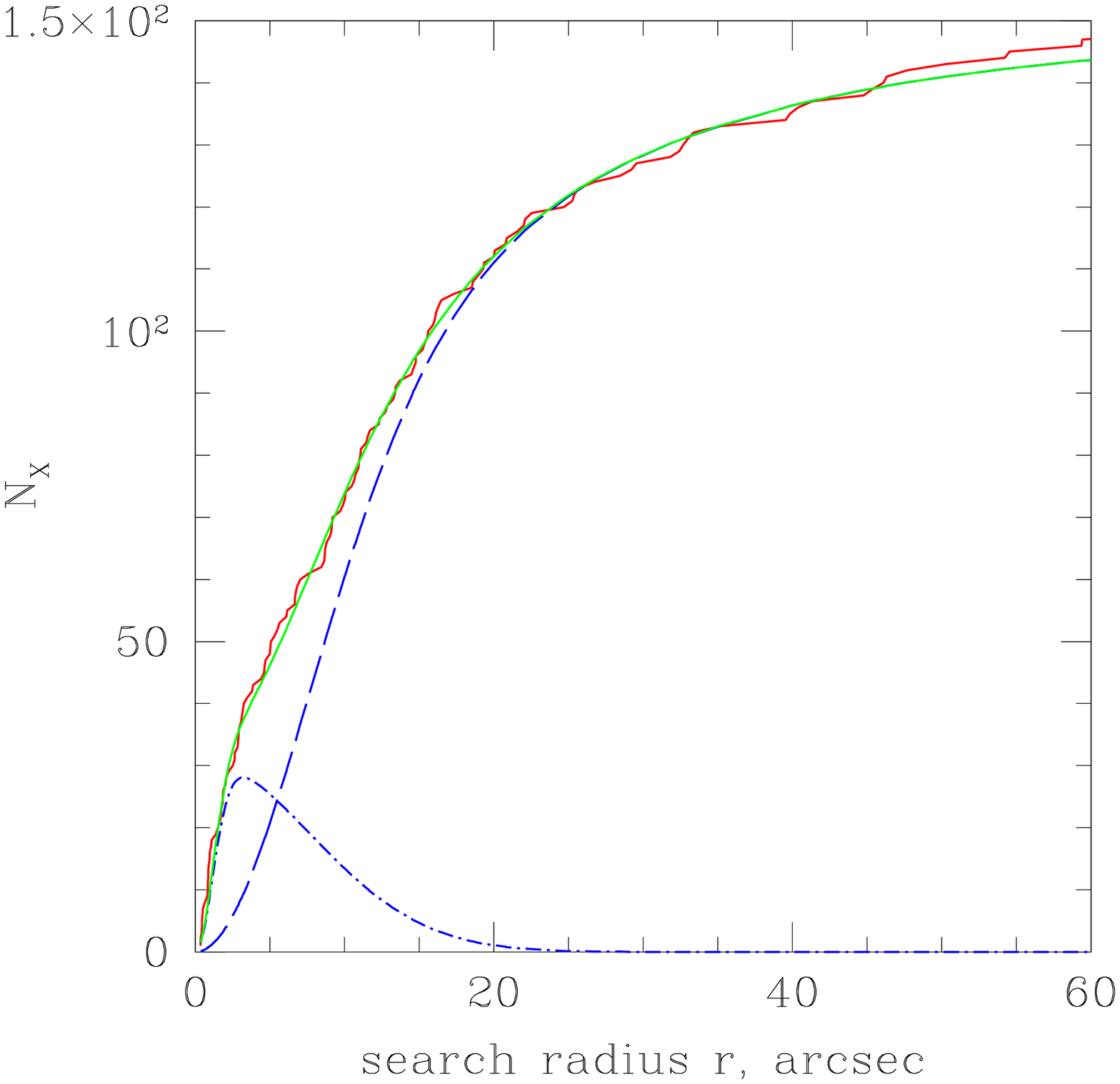}
\caption{The number of X-ray sources with an optical match
as a function of search radius and its 
approximation with eq.~\ref{eq:appendix3}.
The individual components of this equation
corresponding to spurious matches with field stars and true optical 
counterparts (which don't have simultaneous spurious matches)
 are shown with dashed and dot-dashed lines respectively.
}
\label{fig:nmatch2}
\end{figure}

\begin{figure}
\includegraphics[width=0.45\textwidth]{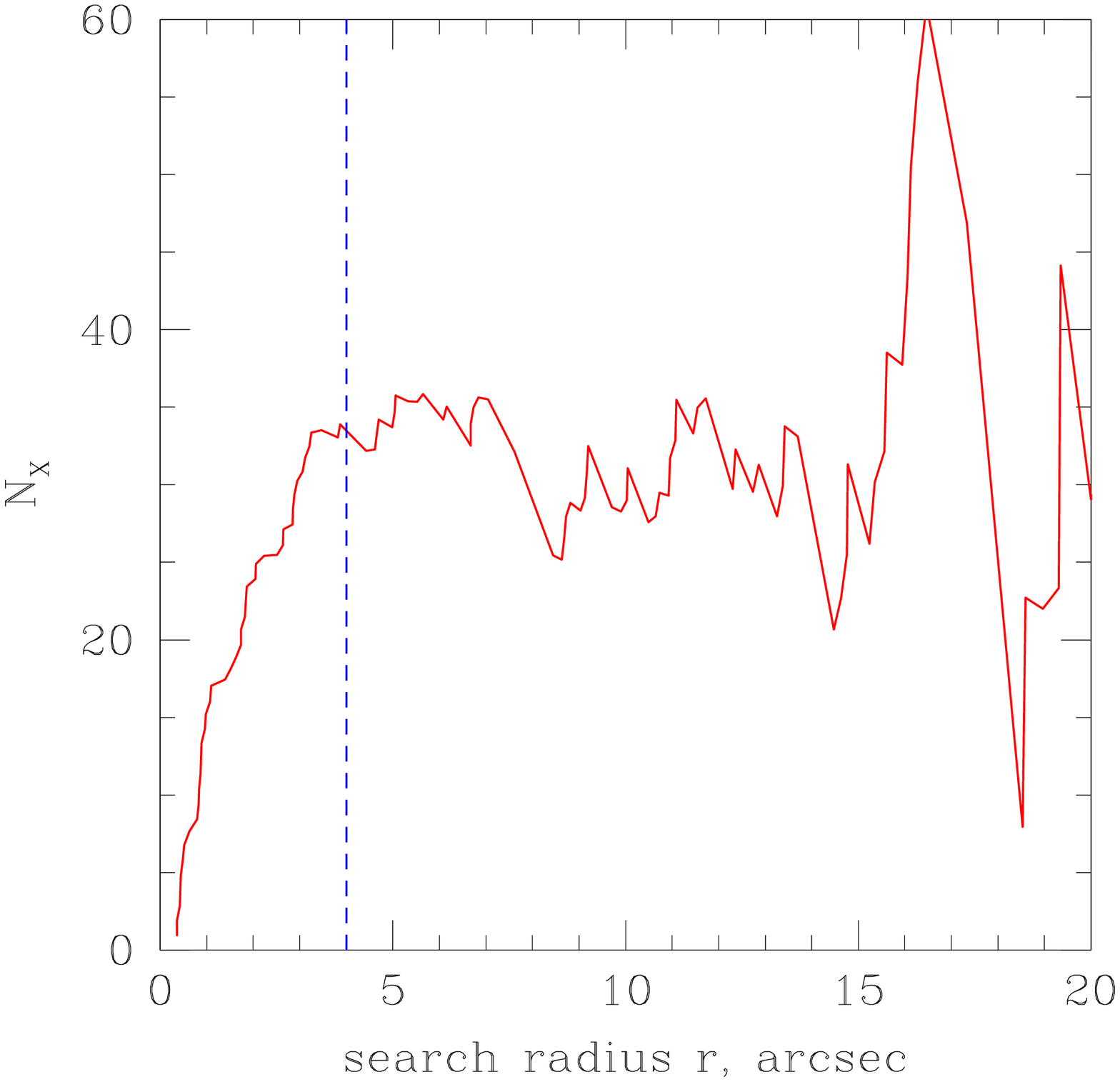}
\caption{
Number of true matches as a function of radius estimated using
eq.~\ref{eq:appendix3} (see ~\ref{sec:appvardens}). The vertical dashed line shows the
search radius of $4$ arcsec.
It is clear from this figure, that almost all true optical counterparts of 
X-ray sources fall within the chosen search radius.}
\label{fig:nmatch}
\end{figure}

Fit parameters are the number of X-ray sources with true optical matches M, the
value of positional error $\sigma$ and the mean density of optical objects $\rho$. 
Fitting the N$_{Xmatch}$(r) curve using real list of XMM sources 
and optical sources from the MCPS
catalogue we obtain the best fit value for the total number of 
sources with true optical
counterpart in our sample is $\sim31$, surprisingly close to the
number of our likely HMXB candidates (see Fig.~\ref{fig:nmatch2}).
Best fit value of positional error, $\sigma\approx1.2\arcsec$, also agrees with
typical values for X-ray sources from our list ranging from $\approx0\arcsec$ to
$\approx2\arcsec$.
We note also, that from the eq.~\ref{eq:appendix3} it follows that subtracting 
the first sum from $N_{Xmatch}$ curve and multiplying the remaining part by $e^{\rho\pi r^2}$,
 we can obtain the dependence of the number of true counterparts among optical matches
 on the search radius.
The resulting curve is shown on Fig.~\ref{fig:nmatch}. 
It is clear from this figure, that almost all true optical counterparts of 
X-ray sources will be detected with the search radius of 4$\arcsec$ 
used in our HMXB identification procedure.

\subsection{Account for spread in distribution of positional uncertainties}

The deviation of positional errors distribution from the delta function
together with the astrometric uncertainty can also modify  N$_{Xmatch}$(r) curve.
With the spread in positional errors accounted for, the equation~\ref{eq:appendix3} becomes: 

\begin{eqnarray}
N_{Xmatch}=\sum_i(1-e^{-\rho_i\pi r^2})+Me^{-\rho\pi r^2}\cdot
\label{eq:appendix4}
\\
~~~~~~~~~~~~~~~~~~~~~~~\cdot\int\int\phi(r',\sigma)g(\sigma)d\sigma dr',
\nonumber
\end{eqnarray}
where $\phi(r,\sigma)\propto re^{-r^2/2\sigma^2}$ is the (Gaussian) distribution
of distances between the X-ray and optical sources for given $\sigma$
and $g(\sigma)$ is a distribution of positional uncertainties of X-ray sources.
To show that such a spread will not significantly change our results, we 
assumed that errors distribution follows the power law
$g(\sigma)\propto\sigma^{\alpha}$ between $\sigma=0.2\arcsec$ 
and $2.3\arcsec$.
In this case the free fit parameters are the number of X-ray sources with true
counterparts M, optical sources density $\rho$ and slope $\alpha$.

Fitting our curve with this formula we obtained the best fit value M=34, which also
agrees well with the number of our likely HMXB candidates.

\label{lastpage}


\begin{thebibliography}{}

\bibitem[\protect\citeauthoryear{Bell \& de Jong}{2001}]{m2l}
Bell E., de Jong R., 2001, ApJ, 550, 212

\bibitem[\protect\citeauthoryear{Bell}{2003}]{bell03}
Bell E., 2003, ApJ, 586, 794


\bibitem[\protect\citeauthoryear{Coe et al.}{Coe et
al.}{2004}]{hmxbopt}  M.J. Coe, W.R.T. Edge, J.L. Galache, 
V.A. McBride, 2004, astro-ph/0410074


\bibitem[\protect\citeauthoryear{Corbet}{Corbet}{1986}]{corbet}
Corbet R.H.D., 1986, MNRAS, 220, 1047  



\bibitem[\protect\citeauthoryear{Cutri et al.}{Cutri et
al.}{2003}]{2mass} Cutri R. M., Skrutskie M. F., van Dyk S.,
Beichman C. A., Carpenter J. M., Chester T., Cambresy L. et al., 2003, yCat, 224



\bibitem[\protect\citeauthoryear{de Vaucouleurs et al.}{de Vaucouleurs
et al.} {1991}]{rc3} de Vaucouleurs G. et al., 1991, Third Refererence
Catalog of Bright Galaxies. Springer-Verlag (RC3) 


\bibitem[\protect\citeauthoryear{Edge \& Coe}{2003}]{edgecoe}	
Edge W. R. T.,  Coe M. J., 2003, MNRAS, 338, 428

\bibitem[\protect\citeauthoryear{Edge et al.}{2004}]{edge}	
Edge W. R. T.,  Coe M. J., Galache J. L., McBride V. A., Corbet
R. H. D., Markwardt C. B., Laycock S., 2004, MNRAS, 353, 1286

\bibitem[\protect\citeauthoryear{Evans et al.}{2004}]{2df} 
Evans C.J., Howarth I.D., Irwin M.J., Burnley
A.W., Harries T.J., 2004, MNRAS, 353, 601


\bibitem[\protect\citeauthoryear{Filipovic et al.}{1998}]{filipovic98}
Filipovic M. et al., 1998, A\&A Auppl., 127, 119

\bibitem[\protect\citeauthoryear{Filipovic et al.}{2000}]{filipovic00}
Filipovic M. D., Pietsch W., Haberl F., 2000, A\&A, 361, 823

\bibitem[\protect\citeauthoryear{Gehrz et al.}{Gehrz et al.}{1974}]{beinfrared}
Gehrz R. D., Hackwell J. A., Jones T. W., 1974, ApJ, 191, 675


\bibitem[\protect\citeauthoryear{Gilfanov}{Gilfanov}{2004}]{gilfanov04}
Gilfanov M., 2004, MNRAS, 349, 146


\bibitem[\protect\citeauthoryear{Grimm, Gilfanov \& Sunyaev}{Grimm et
al.}{2003}]{grimm03} Grimm, H.-J., Gilfanov, M.R.,
Sunyaev, R.A. 2003, MNRAS, 339, 793


\bibitem[\protect\citeauthoryear{Haberl \& Pietsch}{2004}]{BeXraySMC}
Haberl F., Pietsch W., 2004, A\&A, 414, 667

\bibitem[\protect\citeauthoryear{Haberl et al.}{2004}]{haberllongperiod}
Haberl F., Pietsch W., Schartel N., Rodriguez P., Corbet
R. H. D., 2004, A\&A, 420L, 19

\bibitem[\protect\citeauthoryear{Harris \& Zaritsky}{2004}]{sfhistory}
Harris J., Zaritsky D., 2004, AJ, 127, 1531

\bibitem[\protect\citeauthoryear{Helou et al.}{1985}]{helou85}
Helou et al., 1985, ApJ, 298, L7


\bibitem[\protect\citeauthoryear{Kennicutt}{Kennicutt}{1991}]{kenn91}
Kennicutt R,C. Jr. 1991, in: Haynes R.F. \& Milne D.K. (eds.)
Proc. IAU Symp. 148, The Magellanic Clouds, Reidel, Dordrecht, p.139
 
\bibitem[\protect\citeauthoryear{Kennicutt}{Kennicutt}{1998}]{kenn98}
Kennicutt R.C. Jr., 1998, ARA\&A, 36, 189  


\bibitem[\protect\citeauthoryear{Kennicutt et al.}{Kennicutt et
al.}{1995}]{kenn95} Kennicutt R.C. Jr, Bresolin F.,
Bomans D.J., Bothun G.D., Thompson I.B., 1995, AJ, 109, 594  




\bibitem[\protect\citeauthoryear{Lamb et al.}{Lamb et
al.}{2002}]{axp} Lamb R. C., Fox D. W., Macomb D. J., Prince
T. A., 2002, ApJ, 574, 29

\bibitem[\protect\citeauthoryear{Majid et al.}{2004}]{majid}
Majid W. A., Lamb R. C., Macomb D. J., 2004, ApJ, 609, 133 


\bibitem[\protect\citeauthoryear{Massey et al.}{Massey et
al.}{2002}]{massey} Massey P., 2002, yCat, 2236, 0



\bibitem[\protect\citeauthoryear{McGlynn, Scollick \& White}{McGlynn,
Scollick \& White}{1996}]{skyview} McGlynn T., Scollick K., White N., SkyView:
     The Multi-Wavelength Sky on the Internet, McLean, B.J.
     et al., New Horizons from Multi-Wavelength
     Sky Surveys, Kluwer Academic Publishers, 1996,
     IAU Symposium No. 179, p465.
     

\bibitem[\protect\citeauthoryear{Meyssonnier et al.}{Meyssonnier et al.}{1993}]{meyssonnier} 
Meyssonnier N., Azzopardi M., 1993, A\&AS, 102, 451


\bibitem[\protect\citeauthoryear{Monet et al.}{Monet et
al}{2003}]{usno-b} Monet D.G., Levine S.E., Canzian B., Ables H.D.,
Bird A.R., Dahn C.C., Guetter H.H. et al., 2003, AJ, 125, 984

\bibitem[\protect\citeauthoryear{Moretti et al.}{Moretti et
al}{2003}]{cxb} Moretti A., Campana S., Lazzati D., Tagliaferri
G., 2003, ApJ, 588, 696 
 
\bibitem[\protect\citeauthoryear{Morrison \& McLean}{Morrison \&
McLean}{2001}]{gsc} Morrison J. E., McLean B., GSC-Catalog
Construction Team, II, 2001, DDA, 32.0603


\bibitem[\protect\citeauthoryear{Negueruela}{Negueruela}{1998}]{neguer} 
Negueruela I., 1998, A\&A, 338, 505


\bibitem[\protect\citeauthoryear{Shtykovskiy \& Gilfanov}{2005}]{lmcpaper}
Shtykovskiy P., Gilfanov M., 2005, A\&A, 431, 597

\bibitem[\protect\citeauthoryear{Silverman}{1981}]{silverman} 
Silverman B.W., 1981, J.R.Statist.Soc.B, 43, 97

\bibitem[\protect\citeauthoryear{Udalski et al.}{1998}]{ogle}
Udalski A., Szymanski M., Kubiak M. et al., 1998, yCatp005004802U 

\bibitem[\protect\citeauthoryear{van Paradijs \&
McClintock}{1994}]{jvp95} van Paradijs, J. \& McClintock J.E., 1995, 
X-ray Binaries, Cambridge Univ.Press, p.58

\bibitem[\protect\citeauthoryear{Vangioni-Flam et al.}{1980}]{uv80}
Vangioni-Flam E. et al., 1980, A\&A, 90, 73


\bibitem[\protect\citeauthoryear{Westerlund}{Westerlund}{1997}]{mc_book97}  
Westerlund B., ``The Magellanic Clouds'', Cambridge Univ.Press, 1997

\bibitem[\protect\citeauthoryear{Wilke et al.}{2004}]{wilke04}  
Wilke K., Klaas U., Lemke D. et al., 2004, A\&A, 414, 69


\bibitem[\protect\citeauthoryear{Yokogawa et al.}{2000}]{asca} 
Yokogawa J., Imanishi K., Tsujimoto M., Nishiuchi M., 
Koyama K., Nagase F., Corbet R.H.D., 2000,
ApJS, 128, 491

\bibitem[\protect\citeauthoryear{Yokogawa et al.}{2003}]{yoko03} 
Yokogawa J., Imanishi K., Tsujimoto M., Koyama
K., Nishiuchi M., 2003, PASJ, 55, 161


\bibitem[\protect\citeauthoryear{Zaritsky et al.}{Zaritsky et
al.}{2002}]{mcps} Zaritsky D., Harris J., Thompson I.B.,
Grebel E.K., Massey P., 2002, AJ, 123, 855


\end{thebibliography}
\end{document}